\documentclass[a4,10pt,twoside]{article}

\usepackage[utf8]{inputenc}
\usepackage[english]{babel}
\usepackage[T1]{fontenc}
\usepackage{url}
\usepackage[colorlinks=true, allcolors=blue]{hyperref}
\usepackage{graphicx}
\usepackage[left=85pt, right=85pt, top=85pt, bottom=95pt,
headheight=15pt,%
headsep=10pt]{geometry}%
\usepackage{amsmath, amssymb}
\usepackage{authblk}
\usepackage{doi}
\usepackage{newtxtext,newtxmath}
\usepackage{microtype}
\usepackage{booktabs}
\usepackage[bf,small]{caption}

\usepackage[figurename=Fig.]{caption}

\usepackage{titlesec}
\titlespacing*{\paragraph}{0pt}{8pt plus 6pt}{0.5em}

\title{Accuracy assessment of scalar wave propagation methods for diffractive optics design:
  from thin elements to thick binary gratings}

\author[1,*]{Nicolas Barré}

\affil[1]{Independent researcher}

\affil[*]{email: nicolas.barre@protonmail.com}

\date{\today}

\begin{document}

\maketitle

\begin{abstract}
  We present a systematic accuracy assessment of the thin-element approximation (TEA), the
  beam propagation method (BPM), and the wave propagation method (WPM) for binary
  diffractive gratings, using the rigorous Fourier modal method (FMM) as a reference. Random
  binary gratings are generated over a range of spatial frequency cutoffs and thicknesses,
  and the transmitted field overlap between each scalar method and the reference is
  measured.  The results are summarized as accuracy maps in the spatial frequency--thickness
  parameter space, revealing the domain of validity of each method and providing practical
  guidelines for the choice of forward model in diffractive optics inverse design pipelines.
\end{abstract}

\section{Introduction}

The design of diffractive optical elements (DOEs) increasingly relies on gradient-based
inverse design methods, in which a forward propagation model is differentiated with respect
to structural parameters to minimize a figure of merit. The accuracy of such designs is
fundamentally limited by the accuracy of the forward model: an insufficiently accurate
propagation method may converge to a design that performs well in simulation but poorly in
practice. Conversely, unnecessarily rigorous methods incur prohibitive computational costs
that preclude iterative optimization over large parameter spaces.
 
Scalar wave propagation methods occupy a natural middle ground between the thin-element
approximation (TEA) and fully rigorous methods such as finite-difference time-domain (FDTD)
or Fourier modal methods (FMM). TEA reduces the grating to an infinitely thin phase mask,
neglecting any propagation effects within the structure.While computationally trivial, TEA
becomes increasingly inaccurate as the grating thickness grows beyond a few wavelengths, or
when large diffraction angles are involved, since it neglects both the propagation phase
accumulated within the structure and the coupling between spatial frequencies at
interfaces. At the other extreme, FDTD and FMM provide rigorous solutions to Maxwell's
equations but scale poorly with system size, making them impractical as forward models in
iterative optimization loops. The beam propagation method (BPM) and the wave propagation
method (WPM)~\cite{Schmidt:16} offer intermediate accuracy at moderate computational
cost. WPM was introduced to overcome the paraxial limitation of BPM, supporting wide-angle
propagation through the use of the exact dispersion relation $k_z = \sqrt{k^2 - k_\perp^2}$
rather than its paraxial approximation. This makes WPM particularly attractive as a forward
model for the inverse design of thick diffractive structures, where paraxial errors
accumulate over many propagation steps.
 
Despite the widespread use of these methods, a systematic quantitative assessment of their
accuracy as a function of grating parameters is lacking in the literature. Such an
assessment is practically useful for informing the choice of forward model in inverse design
pipelines, and for identifying the regimes where more rigorous methods are necessary.
 
In this work, we present a systematic accuracy assessment of TEA, BPM, and WPM for binary
diffractive gratings, using the rigorous Fourier modal method as a
reference~\cite{schubert2023fourier}. We generate ensembles of random binary gratings
parameterized by their spatial frequency cutoff $f_c$ and physical thickness $h$, and
measure the transmitted field overlap between each scalar method and the rigorous
reference. The results are presented as accuracy maps in the $(f_c, h)$ plane, revealing the
domain of validity of each method and quantifying the conditions under which WPM provides a
significant advantage over BPM. Binary gratings are chosen as a conservative benchmark,
since their sharp index discontinuities represent a more demanding test case than smooth
continuous-relief structures. The scalar propagation methods are implemented and evaluated
using FluxOptics.jl~\cite{Barre:2026:FluxOptics}, a GPU-accelerated differentiable wave
propagation framework implemented in Julia.

\section{Methods}

\paragraph{Grating generation.}
All gratings are defined on a square periodic cell of side $L = 7.5\,\mu$m, discretized on a
$300 \times 300$ grid with pixel size $\Delta x = \Delta y = 25\,$nm~$\approx \lambda/21$,
where $\lambda = 532\,$nm. The refractive indices are $n_1 = 1.5$ (glass substrate) and
$n_2 = 1.0$ (air). Each grating is generated by drawing a uniform random height distribution
over $[0, h]$, applying a circular low-pass filter of cutoff frequency $f_c$ in Fourier
space, rescaling the result to $[0, h]$, and binarizing to $\{0, h\}$ by thresholding at
$h/2$. Filtering is performed via FFT on the periodic grid, so the resulting structure is
exactly periodic with period $L$ by construction. The cutoff frequency $f_c$ is sampled at
11 equally spaced values from $0$ to $f_{c_\mathrm{max}} = 1.33\,\mu\mathrm{m}^{-1}$,
corresponding to a maximum diffraction angle
$\theta_{\max} = \arcsin(\lambda f_c) \approx 45^\circ$ in air. The grating thickness $h$ is
sampled at 11 equally spaced values from $\lambda$ to $11\lambda$, corresponding to phase
delays from $\pi$ to $11\pi$ in the thin-element approximation, since $(n_1 - n_2) = 0.5$
for the chosen material pair. For each $(f_c, h)$ pair, 10 independent random realizations
are generated, yielding a total of $11 \times 11 \times 10 = 1210$ gratings.  We note that
thresholding a smoothly varying field near $h/2$ may generate features narrower than
$1/(2f_c)$ at locations where the gradient is small, making the benchmark slightly more
conservative than the nominal cutoff suggests.
 
\paragraph{Propagation methods.}
All three scalar methods are implemented in FluxOptics.jl~\cite{Barre:2026:FluxOptics} and
share the same simulation geometry: a unit-amplitude plane wave at normal incidence,
polarized along $x$, propagates through a padding layer of thickness
$h_\mathrm{pad} = 0.25\,\mu$m in the substrate ($n_1$), then through the grating of
thickness $h$, and finally through a padding layer of thickness $h_\mathrm{pad}$ in air
($n_2$). The transmitted field is evaluated at the exit of the output padding layer. The
padding layers serve to define homogeneous semi-infinite media at the entrance and exit of
the structure, consistent with the boundary conditions required by the reference FMM
computation.

In the TEA, the grating is represented as an infinitely thin phase element with
transmittance $t(x,y) = \exp\!\bigl(i k_0 (n_1 - n_2)\, S(x,y)\bigr)$, where $S(x,y)$ is the
local surface height. The propagation sequence within the grating region is: free-space
propagation over $h$ in the effective medium of index $(n_1+n_2)/2$, followed by application
of the phase mask.  All free-space steps use the angular spectrum method. This ordering was
found empirically to minimize the phase error relative to the rigorous reference.
 
In BPM, the grating volume is sampled into $n_z = \lfloor h / \Delta z \rfloor$ slices of
thickness $\Delta z = \Delta x = 25\,$nm, using a reference index $n_0 = (n_1+n_2)/2$. The
local index perturbation is $\delta n(x,y,z) = (n_1 - n_2)/2$ where $S(x,y) \geq z$ (glass
region) and $\delta n = -(n_1-n_2)/2$ otherwise, so that $n_0 + \delta n$ recovers the local
refractive index exactly.

In WPM~\cite{Schmidt:16}, the total optical path $h + 2h_\mathrm{pad}$ is divided into
$n_\mathrm{slices} = 30$ uniform steps regardless of $h$, so that the step size
$\Delta z = (h + 2h_\mathrm{pad}) / 30$ varies with $h$. At each step, the field is propagated
using the exact non-paraxial propagator $\exp(i k_z \Delta z)$ with $k_z = \sqrt{(n k_0)^2 -
k_\perp^2}$, evaluated independently in each refractive index region. The binary index profile
is represented with a smooth transition layer at each interface to ensure sub-pixel accuracy
in the height representation~\cite{Barre:2025:Freeform}. The number of slices is capped at 30
to prevent the accumulation of field-matching errors at binary index discontinuities, which
otherwise cause energy non-conservation for large $n_\mathrm{slices}$.
 
\paragraph{Reference method.}
The rigorous reference is computed using FMMAX~\cite{schubert2023fourier}, a GPU-accelerated
Fourier modal method implementation based on the scattering-matrix
formalism~\cite{Liu2012}. The scattering-matrix approach is numerically stable for
arbitrarily thick layers, unlike transfer-matrix formulations which suffer from exponential
ill-conditioning due to evanescent modes. The computation is performed in double precision
(float64), which is required for stable eigendecomposition of the large Fourier-space
matrices at high truncation orders. We use the \texttt{JONES\_DIRECT\_FOURIER} vector-field
formulation with 1600 Fourier orders.  Convergence was verified by cross-validation against
FDTD simulations (Meep~\cite{Oskooi2010}) on selected designs at maximum $(f_c, h)$.
 
\paragraph{Accuracy metric.}
For each grating realization, the complex transmitted field $E(x,y)$ computed by each scalar
method is compared to the $x$-polarized component of the FMM reference field
$E_{\mathrm{ref},x}(x,y)$ via the normalized field overlap:
\begin{equation}
    \eta = \frac{\left|\iint E(x,y)\, E_{\mathrm{ref},x}^*(x,y)\,
    \mathrm{d}x\,\mathrm{d}y\right|^2}
    {\iint |E|^2\, \mathrm{d}x\,\mathrm{d}y \cdot
    \iint |E_{\mathrm{ref},x}|^2\, \mathrm{d}x\,\mathrm{d}y},
\end{equation}
which equals unity for identical fields and is invariant to global phase. The incident field
is $x$-polarized; the metric is evaluated on the co-polarized transmitted component only,
consistent with the scalar approximation shared by all three methods. Cross-polarized
transmission, which can reach up to $5\%$ of the transmitted power for the most demanding
configurations, is therefore excluded from the comparison. Reflected fields are similarly
neglected, as none of the scalar methods account for back-propagation. The reported overlap
for each $(f_c, h)$ point is the mean of $\eta$ over the 10 random grating realizations.
 
\section{Results and Discussion}

\begin{figure}[htbp]
  \centering
  \includegraphics{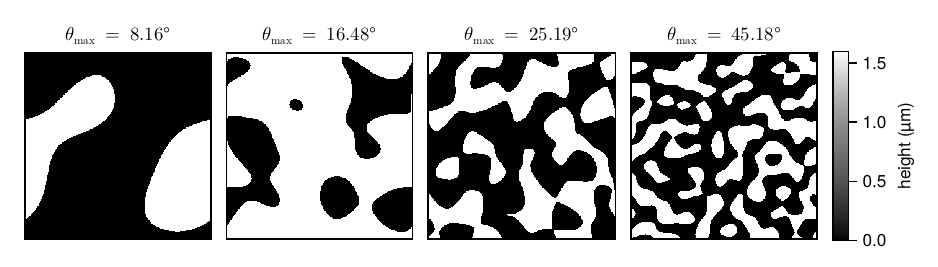}
  \caption{Representative binary grating surface profiles for increasing spatial frequency
    cutoff $f_c$ at fixed thickness $h = 3\lambda$, with $\lambda = 532\,$nm, $n_1 = 1.5$,
    $n_2 = 1.0$. The simulation cell is $L = 7.5\,\mu$m $\times$ $7.5\,\mu$m.
    The colormap encodes the local surface height from $0$ to $h$.}
  \label{fig:gratings}
\end{figure}

Figure~\ref{fig:gratings} illustrates representative binary grating profiles for increasing
spatial frequency cutoff $f_c$ at fixed thickness $h = 3\lambda$. As $f_c$ increases, the
lateral feature size decreases and the grating acquires finer structure, generating larger
diffraction angles. At low $f_c$, the grating consists of smooth, large-scale features that
closely satisfy the TEA assumptions. At high $f_c$, sharp sub-wavelength features dominate
and the interaction of the field with the grating volume becomes increasingly non-trivial.

\begin{figure}[htbp]
  \centering
  \includegraphics{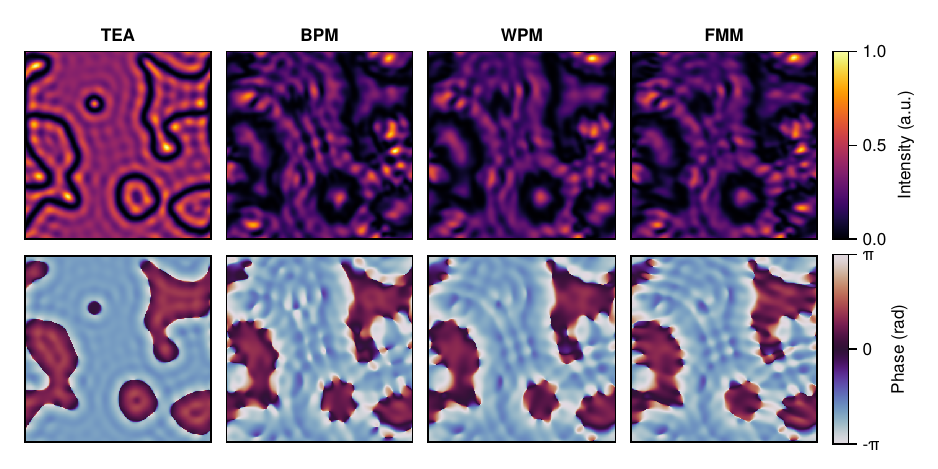}
  \caption{Transmitted intensity (top row) and phase (bottom row) computed by TEA, BPM, WPM,
  and the FMM reference for a representative binary grating at $\theta_{\max} \approx 16.5^\circ$
  and $h = 3\lambda$. Each intensity panel is normalized to its own maximum. The corresponding
  field overlaps with the FMM reference are $\eta_\mathrm{TEA} = 75.1\%$,
  $\eta_\mathrm{BPM} = 89.6\%$, and $\eta_\mathrm{WPM} = 97.3\%$.}
  \label{fig:fields}
\end{figure}

Figure~\ref{fig:fields} shows the transmitted intensity and phase computed by each method
for a representative grating at $\theta_{\max} \approx 16.5^\circ$ and $h = 3\lambda$. The
corresponding field overlaps with the FMM reference are $\eta_\mathrm{WPM} = 97.3\%$,
$\eta_\mathrm{BPM} = 89.6\%$, and $\eta_\mathrm{TEA} = 75.1\%$. WPM reproduces both the
intensity distribution and the phase map of the reference with high fidelity.  BPM shows
visible discrepancies in the intensity and phase profiles, reflecting the accumulation of
errors inherent to the slowly-varying envelope approximation over the propagation
distance. TEA exhibits the largest deviations, which is expected since it entirely neglects
propagation effects within the grating volume.

\begin{figure}[htbp]
  \centering
  \includegraphics{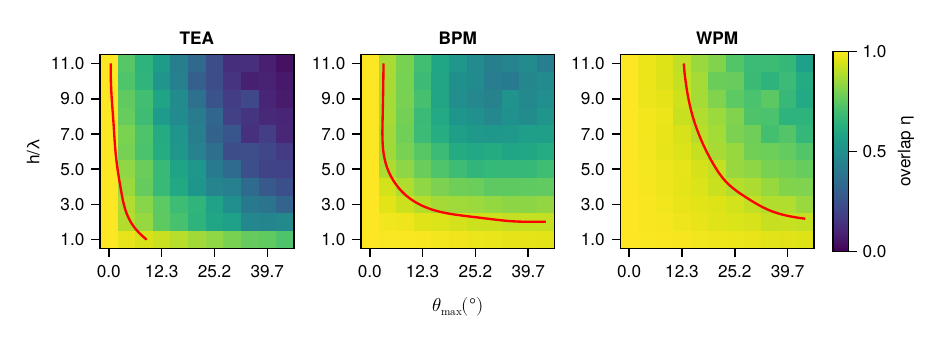}
  \caption{Accuracy maps showing the mean field overlap $\eta$ between each scalar method
  and the FMM reference, as a function of spatial frequency cutoff $f_c$ (expressed as
  maximum diffraction angle $\theta_{\max}$) and grating thickness $h$ (in units of $\lambda$),
  averaged over 10 random binary grating realizations. The red contour indicates the
  $\eta = 0.90$ level. Binary gratings with $n_1 = 1.5$, $n_2 = 1.0$, $\lambda = 532\,$nm.}
  \label{fig:maps}
\end{figure}

The accuracy maps in Figure~\ref{fig:maps} summarize the overlap $\eta$ between each scalar
method and the FMM reference over the full $(f_c, h)$ parameter space, averaged over 10
random grating realizations per point. The red contour indicates the $\eta = 0.90$ level and
serves as a visual guide delimiting the region of acceptable accuracy.

The three methods exhibit markedly different domains of validity. TEA degrades rapidly with
increasing thickness, regardless of $f_c$: already at $h = 3\lambda$ the overlap drops below
$90\%$ for moderate angles, and falls below $50\%$ for $h > 6\lambda$. This confirms that
TEA is fundamentally limited by its inability to account for the phase accumulated during
propagation within the grating volume, a limitation that becomes critical as the grating
thickness increases.
 
BPM performs significantly better than TEA, with the $\eta = 0.90$ contour extending to
larger thicknesses and angles. However, BPM also degrades with $h$, owing to the
accumulation of paraxial phase errors at each propagation step. A notable feature is the
slower decay of BPM at large $h$ compared to TEA, with overlaps stabilizing around
$0.4$--$0.5$ within the explored thickness range.

WPM maintains the highest accuracy across the entire map. For small angles
($\theta_{\max} \lesssim 12^\circ$), $\eta > 0.90$ across the entire explored thickness
range ($h \leq 11\lambda$), with the $\eta = 0.90$ contour remaining nearly vertical,
suggesting that this accuracy could extend to larger thicknesses. As the angle increases,
the accessible thickness for $\eta > 0.90$ decreases progressively, converging toward BPM
performance at $\theta_{\max} \approx 45°$. The advantage of WPM over BPM is therefore most
pronounced for thick structures at moderate angles.  The advantage of WPM over BPM becomes
substantial for thick structures ($h > 3\lambda$) at moderate angles
($\theta_{\max} \lesssim 30°$), where the exact dispersion relation
$k_z = \sqrt{(nk_0)^2 - k_\perp^2}$ prevents the progressive dephasing that limits BPM.
This confirms that WPM is the method of choice for the inverse design of thick diffractive
structures at moderate angles, while BPM remains a competitive alternative for thin gratings
and in the large-angle regime where both methods converge to similar accuracy.

We note that binary gratings represent a conservative benchmark: the sharp index
discontinuities generate a broad spatial frequency spectrum at each propagation step, which
maximally stresses the field-matching procedure at lateral index discontinuities within each
slice. Smooth continuous-relief structures are expected to yield higher overlaps for all
methods at equivalent $(f_c, h)$ parameters, since their gradual index transitions reduce
the spectral content injected at each step. The accuracy maps presented here therefore
constitute lower bounds on the performance of these methods for practical freeform optical
elements. Finally, all three methods share the common limitation of being unidirectional
scalar approximations, neglecting back-propagation and vectorial effects. Extending this
benchmark to the split-step non-paraxial method (SSNP)~\cite{Sharma:04}, which partially
addresses these limitations, is left for future work.

\section{Conclusion}

We have presented a systematic accuracy assessment of three scalar wave propagation methods
(TEA, BPM, and WPM) for binary diffractive gratings, using the rigorous Fourier modal method
as a reference. The resulting accuracy maps in the $(f_c, h)$ plane provide practical
guidelines for the choice of forward model in diffractive optics inverse design: TEA is
adequate only for thin structures ($h \lesssim \lambda$) regardless of angle, BPM extends
the valid regime to moderate thicknesses at low angles or thin structures at large angles,
and WPM is the method of choice for thick structures at moderate angles.

\paragraph{Funding.}
This research received no external funding.

\paragraph{Data availability.}
Scalar propagation simulations were performed using
FluxOptics.jl~\cite{Barre:2026:FluxOptics}, an open-source Julia package for differentiable
wave optics. FMM reference simulations used FMMAX~\cite{schubert2023fourier}. Data and code
underlying the results presented in this paper are available from the author upon reasonable
request.

\let\OLDthebibliography\thebibliography
\renewcommand\thebibliography[1]{
  \OLDthebibliography{#1}
  \setlength{\parskip}{1ex}
  \setlength{\itemsep}{0pt plus 1ex}
}

\bibliographystyle{unsrturl}
\bibliography{fs_wpm}

\end{document}